\documentclass[prb,twocolumn,amsmath,amssymb,floatfix, superscriptaddress]{revtex4-1}

\usepackage{color}

\usepackage{graphicx}
\usepackage[caption = false]{subfig}

\usepackage{bm}

\newcommand{\ket}[1]{\left| #1 \right. \rangle}

\begin{document}
\title{Extracting band structure characteristics of GaSb/InAs core-shell nanowires from thermoelectric properties}
\author{Florinda Vi\~nas}
\affiliation{Division of Solid State Physics and NanoLund, Lund University, Box 118, S-221 00, Lund, Sweden}
\author{H. Q. Xu}
\affiliation{Division of Solid State Physics and NanoLund, Lund University, Box 118, S-221 00, Lund, Sweden}
\affiliation{Beijing Key Laboratory of Quantum Devices, Key Laboratory for the Physics and Chemistry of Nanodevices, and Department of Electronics, Peking University, Beijing 100871, China}
\author{Martin Leijnse}
\affiliation{Division of Solid State Physics and NanoLund, Lund University, Box 118, S-221 00, Lund, Sweden}

\begin{abstract}
Nanowires with a GaSb core and an InAs shell (and the inverted structure) are interesting for studies of electron-hole hybridization and interaction effects due to the bulk broken band-gap alignment at the material interface. We have used eight-band $\mathbf{k\cdot p}$ theory together with the envelope function approximation to calculate the band structure of such nanowires. For a fixed core radius, as a function of shell thickness the band structure changes from metallic (for a thick shell) to semiconducting (for a thin shell) with a gap induced by quantum confinement. For intermediate shell thickness, a different gapped band structure can appear, where the gap is induced by hybridization between the valence band in GaSb and the conduction band in InAs. To establish a relationship between the nanowire band structures and signatures in thermoelectrical measurements, we use the calculated energy dispersions as input to the Boltzmann equation and to ballistic transport equations to study the diffusive limit and the ballistic limit, respectively. Our theoretical results provide a guide for experiments, showing how thermoelectric measurements in a gated setup can be used to distinguish between different types of band gaps, or tune the system into a regime with few electrons and few holes, which can be of interest for studies of exciton physics.
\end{abstract}

\maketitle
\section{Introduction}
\begin{figure}[!]
\includegraphics[width=0.5\textwidth,trim={0 1.5cm 0 3cm},clip]{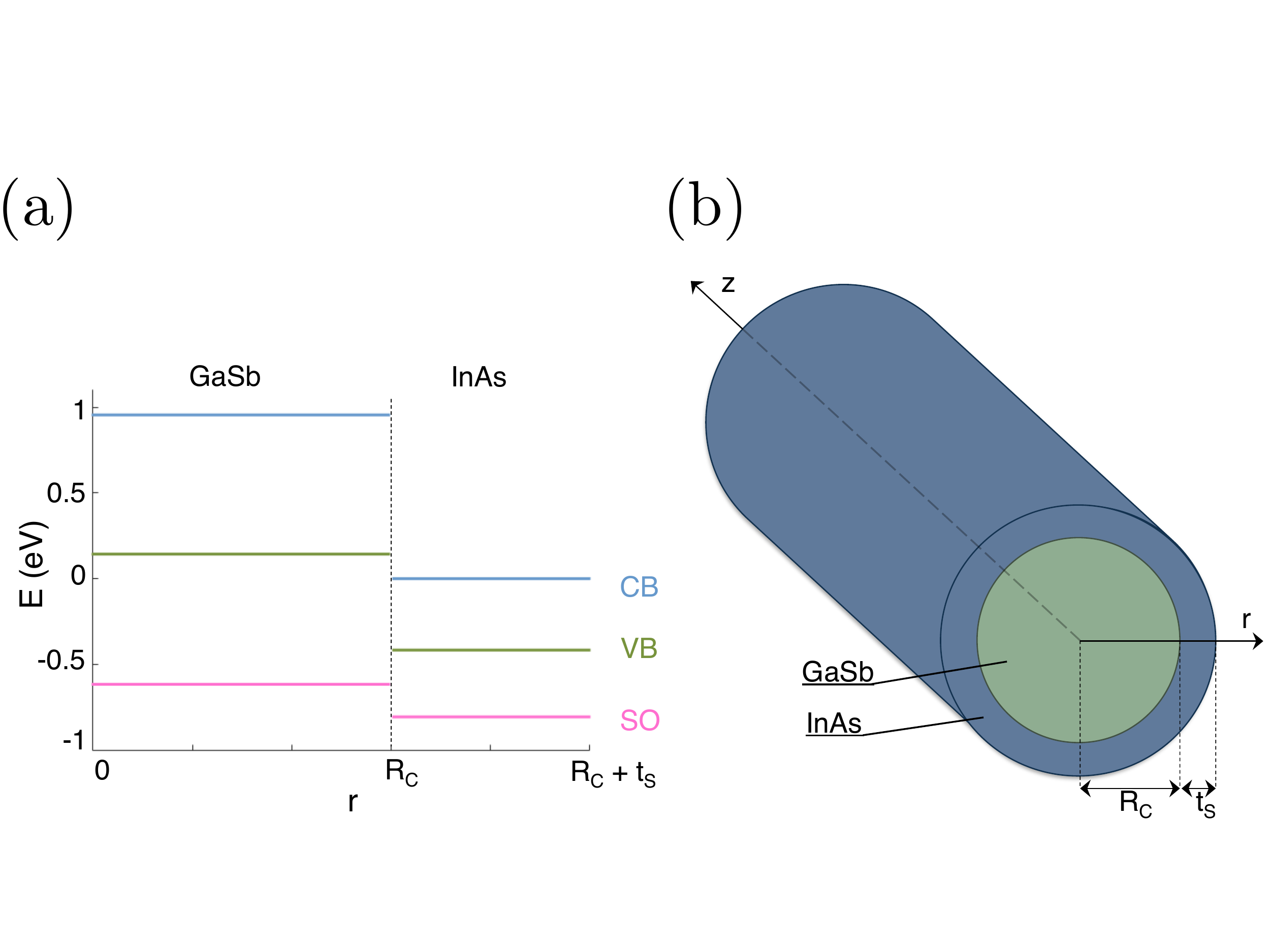}
\includegraphics[width=0.32\textwidth,trim={2cm 6cm 3cm 3cm},clip]{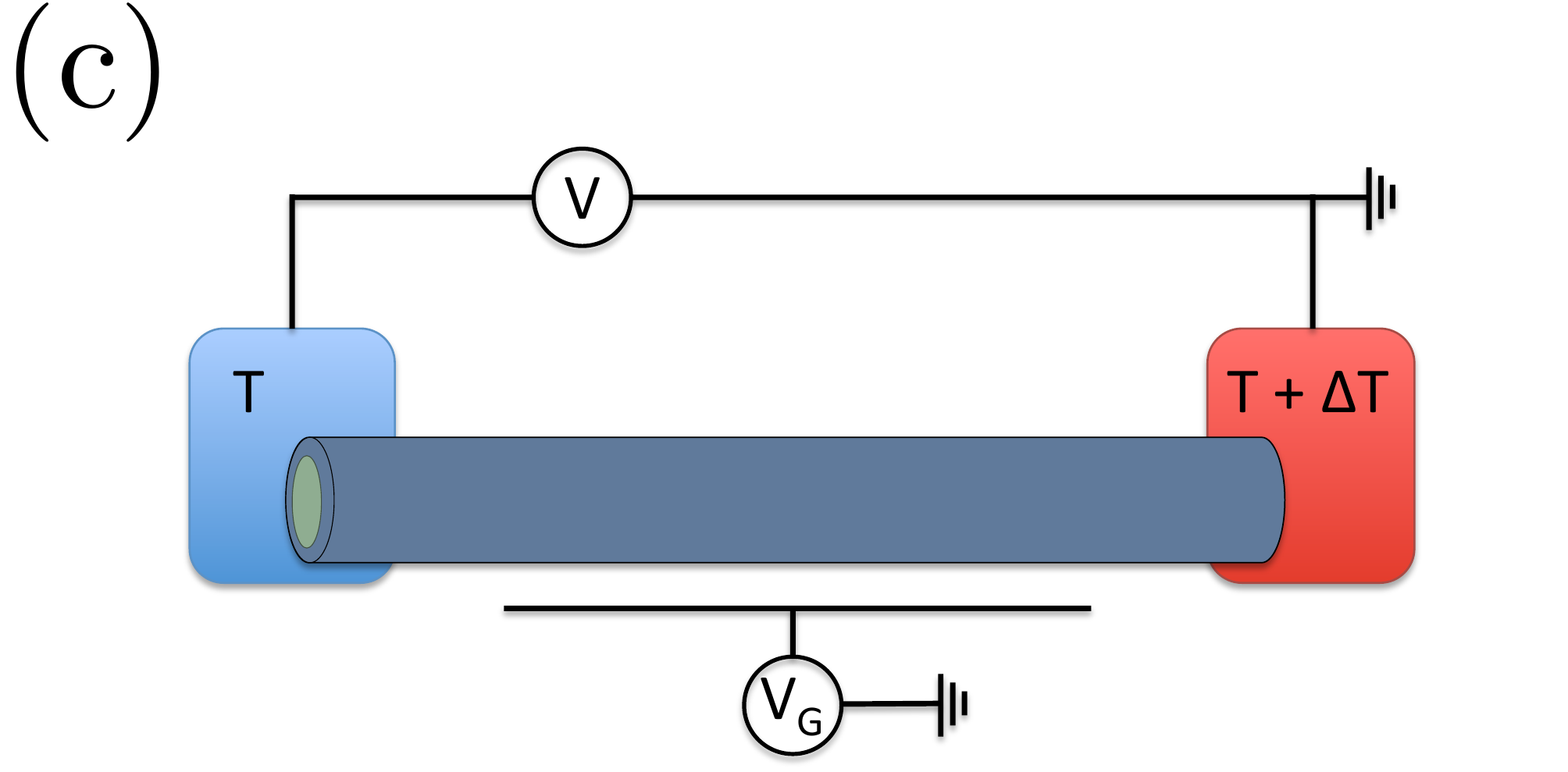}
\caption{a) The GaSb/InAs bulk band alignment with conduction-band edges in blue, valence-band edges in green, and split-off band edges in pink. b) A cylindrical core-shell nanowire with a GaSb core radius $R_C$ and an InAs shell thickness $t_S$. c) A setup for transport measurement of a nanowire. Electrodes are depicted in blue and red, with applied source voltage $V$. The nanowire is subject to a gate voltage $V_G$ and a heat gradient $\Delta T$.}
\label{fig_CS}
\end{figure}

The material system GaSb/InAs is of great interest because it is nearly lattice matched and has a bulk broken band gap alignment; see Fig.~\ref{fig_CS}(a).  The bulk broken band-gap alignment leads to a system with hybridizing electron and hole states close to the interface. When confined in a two-dimensional quantum-well geometry, the hybridization effect can be large enough that the band structure becomes inverted and gapped at the same time; we call such a band gap a hybridization gap. In this case, it has been proven that the GaSb/InAs material system is a topological insulator; it is then the hybridization gap that is responsible for the quantum spin Hall effect.\cite{Liu2008} In two dimensions, the GaSb/InAs material system may also host excitonic ground states\cite{Naveh1996} and interesting spin-orbit effects\cite{Nichele2016}; and in zero dimensions coupled electron and hole quantum dots have been studied.\cite{Ganjipour2015, Nilsson2016}

Recently, it has become possible to grow GaSb/InAs heterostructures in the form of core-shell nanowires.\cite{Ek2011, Rieger2015, Namazi2015} A core-shell nanowire\cite{Lauhon2002} is a heterostructure in the radial direction, see Fig.~\ref{fig_CS}(b), with applications within, e.g., infrared radiation\cite{Mayer2013} and energy harvesting.\cite{Tian2007} The broken band-gap alignment makes GaSb/InAs core-shell nanowires particularly interesting, and recent experiments have shown that the carrier transport can be tuned between hole dominated and electron dominated by using a gate electrode.\cite{Ganjipour2012,Gluschke2015}
Previous theoretical studies have investigated the effect of strain in core-shell nanowires of various material combinations\cite{Pistol2008} and studied the electronic structure of GaSb/InAs core-shell nanowires with strong confinement,\cite{Kishore2012, Luo2016SR} mainly focusing on tuning the wire dimensions to optimize the hybridization gap.

In this paper we present calculations of the band structure and its effect on the carrier transport in GaSb/InAs and InAs/GaSb core-shell nanowires. We focus on conditions that are tractable for epitaxial nanowire growth, meaning a rather thick core radius $R_C$ and growth in the crystallographic $[111]$ direction. 
We calculate the electronic structure for both GaSb/InAs and InAs/GaSb zinc-blende core-shell nanowires of cylindrical cross sections, using eight-band $\mathbf{k\cdot p}$ theory and the envelope function approximation (EFA), to account for confinement, together with a Bessel function basis expansion. 
The choice to approximate the nanowires as cylindrical is motivated by the small changes in energy dispersion that this leads to, in comparison with simulations for hexagonal nanowires.\cite{NWLuo}
In our calculations we focus on nanowires with a fixed core radius $R_C = 20$ nm and show that hybridization between the valence bands (VBs) and the conduction bands (CBs) occurs, and that by choosing suitable dimensions of the shell thicknesses, a hybridization gap opens up for the nanowires with a GaSb core and an InAs shell. However, for core-shell nanowires with an InAs core and a GaSb shell, there is no shell thickness that gives a hybridization gap for such a large core radius. The transport properties of the core-shell nanowires are studied in two limits: the diffusive limit, employing the semiclassical Boltzmann transport equation, and the ballistic limit.  In both of these limits, our calculated energy dispersion is used as input, in a similar fashion to Refs.~\onlinecite{Sofo1994, Scheidemantel2003, May2009, Damico2016}.

Our results indicate that the presence of a hybridization gap can be deduced from transport properties.
Although the conductance looks similar for a nanowire with a hybridization gap and a nanowire with a confinement-induced gap, the two can be distinguished based on the Seebeck coefficient. Similarly, when there is a larger overlap of the GaSb VB and the InAs CB, such that no hybridization gap appears, the thermoelectric transport signatures can reveal the size of this overlap. Therefore, comparing measurements of the Seebeck coefficient with calculated band structures helps to separately determine the electron and  hole densities. 

The remainder of the paper is organized as follows: Section II describes the methods for calculating the band structure and transport properties. Sec. III presents the results. Finally, Section IV summarizes our findings and concludes.

\section{Method}
\subsection{Electronic structure}
In this work, we employ the eight-band $\mathbf{k\cdot p}$ method,\cite{Kane1957} commonly referred to as the Kane model, to obtain the electronic dispersion of the nanowires. The $\mathbf{k\cdot p}$ method is a semiempirical method to solve the Schr\" odinger equation to obtain the band structure. Together with the EFA, it provides a powerful tool for approximative calculations of the electron energies in low-dimensional structures such as quantum wells, nanowires, and quantum dots. For the InAs/GaSb material system it is crucial to use a method such as the eight-band $\mathbf{k\cdot p}$ method, where the interaction between the CB and the VBs is taken into account.

From the Hamiltonian 
\begin{equation}
H = \frac{p^2}{2m_0} + V_0(\mathbf{r}) + \frac{\hbar}{4m_0^2c^2}\mathbf{p\cdot \sigma} \times (\nabla V_0),
\end{equation}
we can obtain the Kane Hamiltonian $H_{8}$, following Refs.~\onlinecite{Kane1957, Winkler2003}. Here, $\mathbf{p}$ is the momentum operator, $m_0$ is the electron rest mass, $V_0$ is the periodic crystal potential, $c$ is the speed of light, and $\sigma$ is the spin operator. We take one s-like CB and three p-like VBs into account explicitly, using the Bloch function basis 
\begin{align}
\begin{split}
\{ &\ket{S \uparrow}, \ket{P_x \uparrow}, \ket{P_y \uparrow}, \ket{P_z \uparrow}, \\
&\ket{S \downarrow},\ket{P_x \downarrow}, \ket{P_y \downarrow}, \ket{P_z \downarrow}   \},
\end{split}
\end{align}
where $S$ and $P$ refer to the symmetry of the CB and the VBs, respectively,\cite{Kane1957} and treat coupling to all other bands perturbatively.
The $8 \times 8$ Kane Hamiltonian can be written as\cite{Foreman1997, Birner2011} $H_{8} = H_0 + H_{SO}$, with
\begin{equation}
H_0 = \left(\begin{array}{c | c}
\begin{array}{c c}
H_{cc}	&	H_{cv}	\\
H_{vc}	&	H_{vv}	
\end{array} & 0 \\
\hline
0 & 	\begin{array}{c c} 
H_{cc}		&	H_{cv}	\\
H_{vc}		&	H_{vv}
\end{array} \\
\end{array}\right) \label{HwoSO}
\end{equation}
and
\begin{equation}
H_{SO} = \frac{1}{3}\Delta_{SO}\begin{pmatrix}
0	&	0	&	0	&	0	&	0	&	0	&	0	&	0 \\
0	&	0	&	-i	&	0	&	0	&	0	&	0	&	1 \\
0	&	i	&	0	&	0	&	0	&	0	&	0	&	-i \\
0	&	0	&	0	&	0	&	0	&	-1	&	i	&	0 \\
0	&	0	&	0	&	0	&	0	&	0	&	0	&	0 \\
0	&	0	&	0	&	-1	&	0	&	0	&	i	&	0 \\
0	&	0	&	0	&	-i	&	0	&	-i	&	0	&	0 \\
0	&	1	&	i	&	0	&	0	&	0	&	0	&	0
\end{pmatrix}.
\end{equation}
Following the Burt-Foreman formalism,\cite{Foreman1997} the submatrices of $H_0$ can be expressed as
\begin{align}
H_{cc} &= E_c + k_xAk_x + k_yAk_y + k_zAk_z, \\
H_{cv} &= \begin{pmatrix}
k_yBk_z + iPk_x & k_zBk_x+iPk_y	&	k_xBk_y + iPk_z \label{Hcv}
\end{pmatrix}, \\
H_{vc}  &= H_{cv}^{\dagger} \label{Hvc}
\end{align}
and
\begin{widetext}
\begin{align}
H_{vv} = &\left( E_v' + \frac{\hbar^2}{2m_0} \mathbf{k}^2 \right)
 \begin{pmatrix}
1	&	0	&	0	\\
0	&	1	&	0	\\
0	&	0	&	1
\end{pmatrix} \\
&+ \begin{pmatrix}
k_xLk_x + k_yMk_y + k_zMk_z		&		k_xN^+k_y + k_yN^-k_x	&	k_xN^+k_z + k_zN^-k_x	\\
\dagger			&	k_xMk_x + k_yLk_y + k_zMk_z	&	k_yN^+k_z + k_zN^-k_y	\\
\dagger			&		\dagger	&	k_xMk_x + k_yMk_y + k_zLk_z
\end{pmatrix}
\end{align}
\end{widetext}
where $E_v' = E_v - \Delta_{SO}/3$.
All parameter definitions can be found in Appendix~\ref{app:kp}, and their values are taken from Ref.~\onlinecite{Vurgaftman2001}, except $P$ that is calculated according to the formula given by Ref.~\onlinecite{Foreman1997} to avoid spurious solutions. All parameters are assumed to vary step-like at the material interface. 
We use $N^+ = N^- = N/2$; for a motivation, see Appendix~\ref{app:kp}. However, it is worth noting that the ordering of the $k$ operators and the parameter $P$ matters, and that the correct version is indeed given by Eqs.~(\ref{Hcv}) and~(\ref{Hvc}), as suggested by Ref.~\onlinecite{Foreman1997}. We assume, like Refs.~\onlinecite{Foreman1997, Lassen2006, Veprek2007, Luo2016SR}, that the parameter $B$ in the Kane model can be set to zero when considering a zinc-blende structure without any external fields.

Choosing $z$ as the nanowire growth direction, we will have 
\begin{equation}
k_{r_i} \rightarrow -i\frac{\partial}{\partial r_i}, \ r_i=x,y
\end{equation}
while $k_z$ is the crystal momentum in the growth direction. 
We solve the Schr\"odinger equation $H_8 \psi = E \psi$ using basis function expansion of the envelope functions $\psi(r,\theta,z)$, in a similar fashion to Ref.~\onlinecite{Gershoni1993}, taking advantage of the rotational symmetry of the cylindrical wire. The expansion we use is given by
\begin{equation}
\psi (r,\theta,z) = \sum_{n = -N}^{N\rightarrow \infty} \sum_{l = 1}^{L\rightarrow \infty} N(n,l) J_n(\alpha_{n l }\frac{r}{R}) e^ {in\theta} e^ {i k_z z}
\end{equation} 
with $J_n$ the first kind Bessel function of order $n$, $\alpha_{nl}$ its zero of order $l$, and $N(n,l)$ a normalization factor. The Bessel functions are scaled by the total nanowire radius $R$. 

To be able to make predictions for nanowires grown in the [111] crystallographic direction we must rotate our system. The rotation process is described in detail in Appendix~\ref{app:rot} and follows Refs.~\onlinecite{Los1996, Lassen2006, Willatzen2009, NWLuo}. In addition to the rotation, we follow Refs.~\onlinecite{Lassen2006,NWLuo}  and make a change of basis after the rotation of the Hamiltonian is performed. The new basis ${\ket{j, j_z	}}$, with the total angular momentum $j$ and the angular momentum in the growth axis $j_z$, diagonalizes the spin-orbit interaction and is also used to label the conduction, heavy hole, light hole and split-off bands (see Fig.~\ref{fig_CS}(a)).

\subsection{Transport calculations}
We now want to calculate electrical transport through the nanowire as a response to the electrical bias $V$, with gradient $\nabla V$, and the temperature difference $\Delta T$, with gradient $\nabla T$, see Fig.~\ref{fig_CS}(c).
We restrict ourselves to the linear response regime, meaning small $V$ ($\nabla V$) and small $\Delta T$ ($\nabla T$), in which case the current density $J$ becomes
\begin{equation}
J = -\sigma \nabla V - \sigma S \nabla T,
 \label{currdens}
\end{equation}
where $\sigma$ is the electrical conductivity and $S$ is the Seebeck coefficient. The corresponding expression for the current is $I = G V + G S \Delta T$, where $G$ is the conductance.
The transport calculations are carried out with the energy dispersions obtained from the $\mathbf{k\cdot p}$ calculations as input. 
We investigate both the limits of diffusive and ballistic transport (we note that ballistic transport has indeed been observed in short nanowire segments\cite{Weperen2013, Chuang2013, Kammhuber2016}).

\subsubsection{Diffusive transport -- the Boltzmann equation}
In the diffusive transport limit, we can find the current density from the distribution function\cite{Lundstrom2002} by solving the Boltzmann equation in the linear response regime. Furthermore, we apply the relaxation time approximation and assume that the relaxation time $\tau$ is the same for all bands and subbands, and that it is independent of $k_z$. We then obtain the following expressions for the conductivity and the Seebeck coefficient:
\begin{align}
 \sigma = 2e^2\tau\sum_{m}\int_{-\infty}^{\infty}dk_z &v_m^2(k_z)\left( -\frac{\partial f(E_m, E_F)}{\partial E} \right), \\
  \sigma S = 2eT\tau\sum_m \int_{-\infty}^{\infty}dk_z &v_m^2(k_z) (E_m-E_F) \nonumber \\
  \times &\left( -\frac{\partial f(E_m, E_F)}{\partial E} \right).
\end{align}
Here, $-e$ is the electron charge, $f(E_m, E_F)$ is the Fermi-Dirac distribution with Fermi level $E_F$, $T$  is the temperature, $\tau$ is the relaxation time, and $E_m$ and $v_m(k_z) = \frac{1}{\hbar} \partial_{k_z} E_m(k_z)$ are the energy and the velocity in band $m$ (by band we mean all one-dimensional subbands seen in Figs.~\ref{GaSbInAs_Hybr} and \ref{InAsGaSb_bands}). 
In general, it is not possible to write these expression as integrals over the energy $E_m$, unless there exists a function $k_z(E_m)$. The integrals over $k_z$ are evaluated numerically with our calculated energy dispersion as input.

\subsubsection{Ballistic transport}

In the ballistic limit\cite{Davies1998} electrons do not thermalize along the nanowire and therefore one cannot define local values or gradients of $V$ and $T$. Instead, the total current is independent of the nanowire length, and depends only on $\Delta V$ and $\Delta T$.
The conductance and the Seebeck coefficient are then given by
\begin{align}
G = \frac{e^2}{2\pi} \sum_{m}\int_{-\infty}^{\infty} &\left( -\frac{\partial f(E_m,E_F)}{\partial E} \right) \nonumber \\
\times & \left(v_m^{\rightarrow}(k_z)- v_m^{\leftarrow}(k_z)\right) T(k_z) dk_z \\
G S  = \frac{e}{2\pi} \sum_{m}\int_{-\infty}^{\infty} &\frac{E_m-E_F}{T}\left( -\frac{\partial f(E_m, E_F)}{\partial E} \right)  \nonumber \\
\times &  \left(v_m^{\rightarrow}(k_z)- v_m^{\leftarrow}(k_z)\right) T(k_z) dk_z.
\end{align} 
We will assume that the potential is even along the entire nanowire such that there is no scattering and the transmission coefficient is $T(k_z)=1$. The velocities $v_{\rightarrow}(k_z)$ and $v_{\leftarrow}(k_z)$ are defined as the carrier velocity for right- and left-moving particles, respectively. 
Here we have used the Kramers degeneracy, $E_+(-k_z) = E_-(k_z)$, so that the first-order term vanishes under the $k_z$ integral when expanding the integrand in a Taylor series. Again, the integration variable $k_z$ cannot be substituted for $E_m(k_z)$, and the integrations must be carried out over $k_z$.

\section{Results}
To find the energy dispersion for the core-shell nanowires, we diagonalize the rotated version of $H_8$ found in Eq.~(\ref{eq_Hrot}). The energies $E_m(k_z)$ are plotted as a function of $k_z$.

	\subsection{GaSb/InAs}
\begin{figure}
\includegraphics[width=0.5\textwidth,trim={0 6.5cm 0.8cm 7cm},clip]{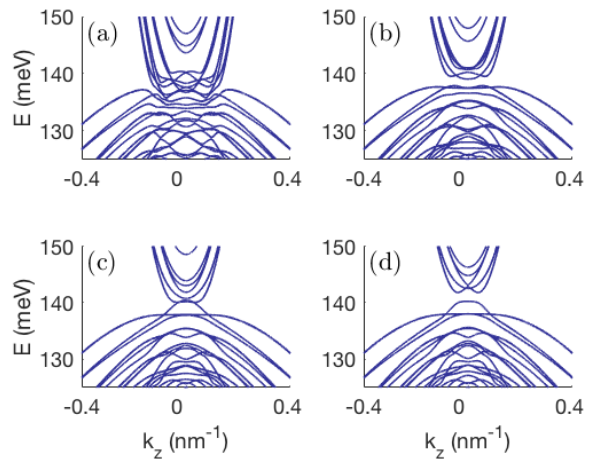}
\caption{Band structure for a GaSb/InAs nanowire with core radius $R_C = 20$ nm and shell thickness a) $t_S = 8.5$ nm, b) $t_S = 7.8$ nm, c) $t_S = 7.62$ nm, and d) $t_S = 7.5$ nm. In b) a hybridization gap of $1.1$ meV opens up. }
\label{GaSbInAs_Hybr}
\end{figure}
Figure~\ref{GaSbInAs_Hybr} shows band structures for GaSb/InAs core-shell nanowires grown in the [111] direction. The core radius is fixed at $R_C =  20$~nm. The band structures at the energies we focus on are dominated by the VB in GaSb and the CB in InAs. Because of hybridization. the bands are far from parabolic, and sometimes vary between positive (CB-like) and negative (VB-like) for different values of $k_z$. In Fig.~\ref{GaSbInAs_Hybr}(a), the bands overlap completely and no band gap is present, so that the system is metallic. A large number of subbands from the quasi-one-dimensional confinement are present, and couplings between bands result in avoided crossings. By decreasing the nanowire shell thickness, a hybridization gap opens up; see Fig.~\ref{GaSbInAs_Hybr}(b). In this case, we have band inversion together with an effective gap. Just below this hybridization gap, a so-called camelback  structure can be seen, where the topmost band has a CB-like appearance around the $\Gamma $ point, but is VB-like for larger $k_z$. In Fig.~\ref{GaSbInAs_Hybr}(c), the shell thickness is decreased further, leading to a closing of the hybridization gap, so that the system is again metallic. For an even thinner shell, a normal confinement band gap opens up, which can be seen in Fig.~\ref{GaSbInAs_Hybr}(d). Both above and below the band gap we see hybridized states, but the camelback feature is not present in the bands closest to the band gap. 
These results are qualitatively similar to those for a [001] wire.\cite{Kishore2012}

\subsection{InAs/GaSb}
\begin{figure}
\includegraphics[width=0.5\textwidth,trim={0.9cm 6cm 1cm 6.4cm},clip]{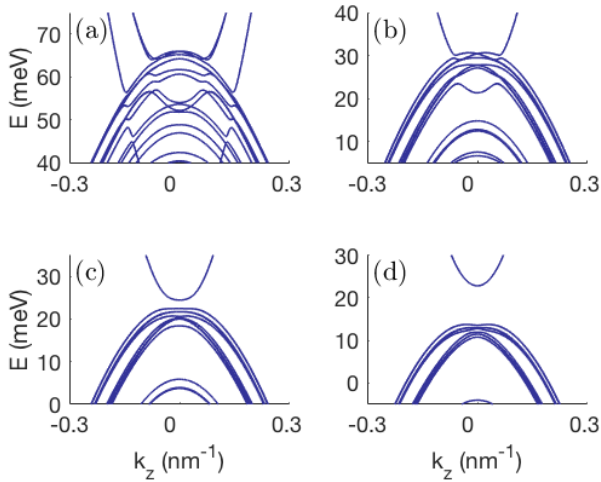}
\caption{Band structure for an InAs/GaSb core-shell nanowire with core radius $R_C = 20$ nm and shell thickness a) $t_S = 2.7 $ nm,  b) $t_S = 2.1$ nm c) $t_S = 2$ nm and d) $t_S = 1.9$ nm. }
\label{InAsGaSb_bands}
\end{figure}
Figure~\ref{InAsGaSb_bands} shows the calculated band structures for InAs/GaSb core-shell nanowires with core radius $R_C = 20$ nm. In Figs.~\ref{InAsGaSb_bands}(a) and~\ref{InAsGaSb_bands}(b), the band structure is ungapped, while in Figs.~\ref{InAsGaSb_bands}(c) and~\ref{InAsGaSb_bands}(d), a band gap is present. 
In contrast to nanowires with GaSb cores, we do not observe a hybridization gap for fixed core radius of $R_C = 20$~nm. 
A clear difference between the nanowires with GaSb and InAs cores is the large difference between the effective hole mass in GaSb ($m_{hh} = 0.71 m_0$) and the effective electron mass in InAs ($m_e = 0.026 m_0$). For a hybridization gap to open up, we need an overlap of CBs and VBs that anticross due to a coupling between them. In the GaSb/InAs material system, most of the holes are confined in the GaSb and the electrons in the InAs. This means that confinement effects are much less pronounced in GaSb than in InAs, so that for nanowires with an InAs core a very thin GaSb shell is needed to see confinement effects in the shell at all. Our calculations show that for such a thin GaSb shell, the hole energies are very sensitive to changes of the shell thickness. This has the consequence that for an InAs/GaSb core-shell nanowire we will have either a confinement gap (for thin shells), or no energy gap at all (for thick shells), but there is no shell thickness such that the nanowire exhibits a hybridization gap. 
However, the band structures for thinner core-shell nanowires (not shown) with InAs cores show qualitatively similar behavior to thinner nanowires with GaSb cores, including the hybridization gap. For core-shell nanowires with a hybridization gap, the size of this gap can be changed by tuning the shell thickness. For an InAs (GaSb) core, a thicker (thinner) shell should be used to maximize the size of the hybridization gap when the core radius is decreased. However, for a GaSb core radius below some critical value, the hybridization gap size will tend to zero, similar to the case for a thick InAs core radius.

\subsection{Transport in core-shell nanowires}

\begin{figure}
\includegraphics[width=0.5\textwidth,trim={1cm 1.5cm 1cm 1.5cm},clip]{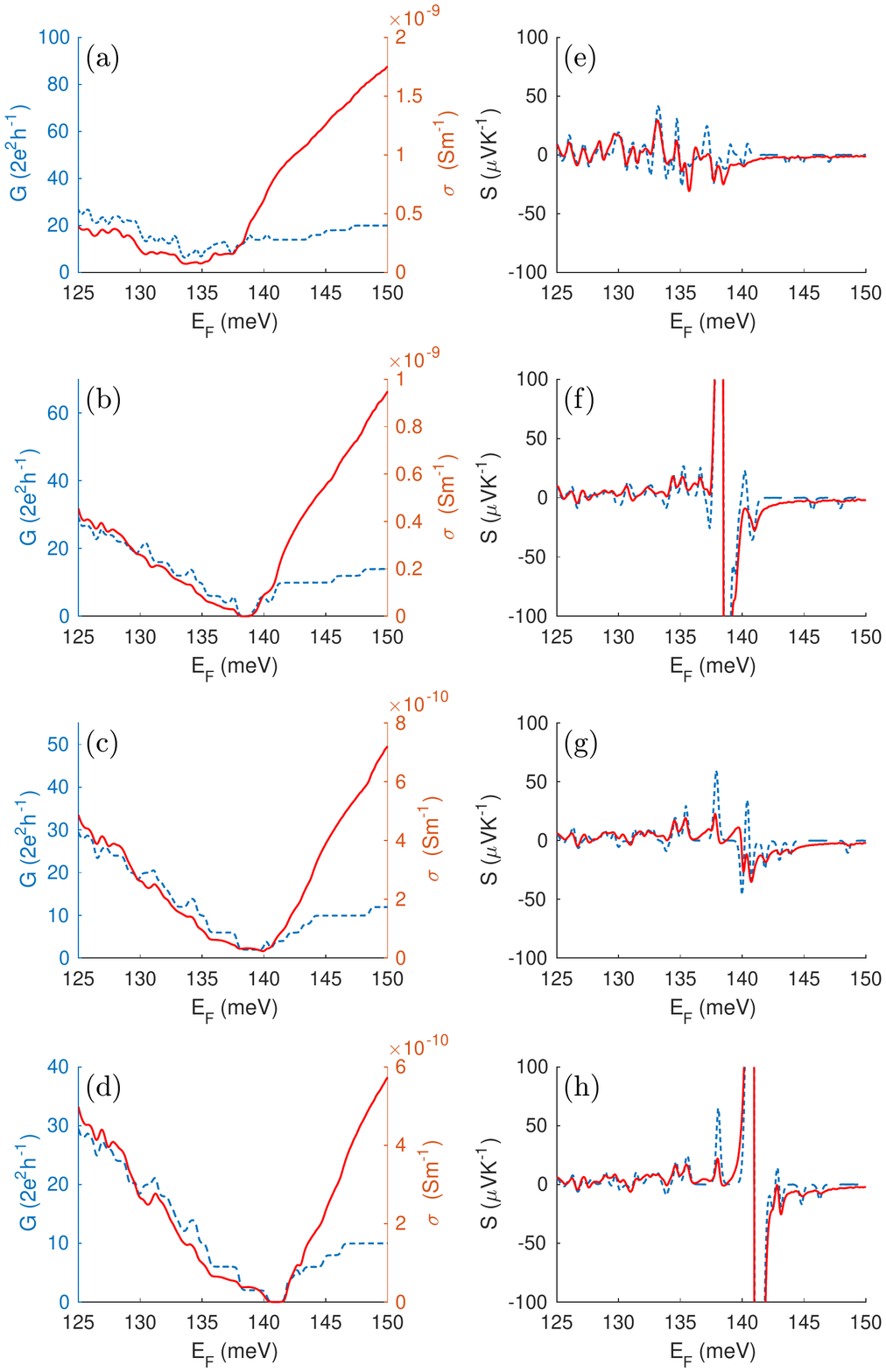}
\caption{(a)--(d) Conductivity (red solid line) and conductance (blue dashed line), calculated in the diffusive and the ballistic limit respectively, for the GaSb/InAs $[111]$-nanowires with core radius $R_C = 20$ nm. 
(e)--(h) Seebeck coefficient, calculated in the diffusive limit (red solid line) and the ballistic limit (blue dashed line), for the same wires. (a)--(d) [(e)--(h)] correspond to the nanowire dimensions in Figs.~\ref{GaSbInAs_Hybr}(a)--(d).
}
\label{trans_GaSb}
\end{figure}

\begin{figure}
\includegraphics[width=0.5\textwidth,trim={1cm 1.5cm 1cm 1.5cm},clip]{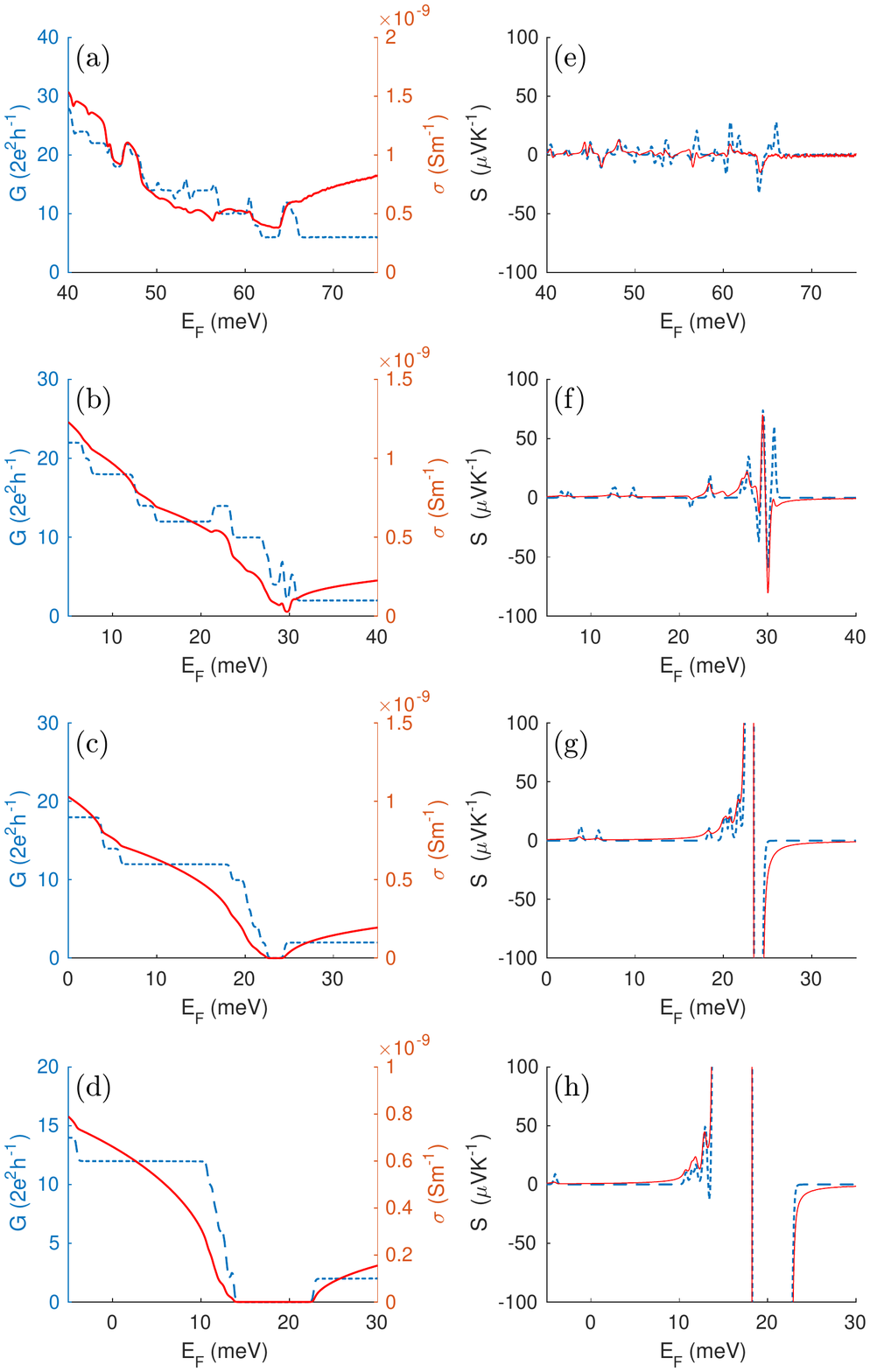}
\caption{(a)--(d)  Conductivity (red solid line) and conductance (blue dashed line), calculated in the diffusive and the ballistic limit respectively, for the InAs/GaSb $[111]$-nanowires with core radius $R_C = 20$ nm. 
(e)--(h) Seebeck coefficient, calculated in the diffusive limit (red solid line) and the ballistic limit (blue dashed line), for the same wires. (a)--(d) [(e)--(h)] correspond to the nanowire dimensions in Figs.~\ref{InAsGaSb_bands}(a)--(d).
}
\label{trans_InAs}
\end{figure}
 
Figure~\ref{trans_GaSb} depicts the calculated conductivities, conductances, and Seebeck coefficients as a function of $E_F$ for the GaSb/InAs core-shell nanowires, corresponding to the band structures in Fig.~\ref{GaSbInAs_Hybr}.\footnote{In the calculations we use $\tau = 0.5$~ps in rough agreement with Refs. \onlinecite{Kash1983} and \onlinecite{Gonzalez1999}. However, it should be noted that $\tau$ only affects the overall magnitude of $\sigma$.} 

A change in $E_F$ can be achieved experimentally by applying a gate voltage. In both the diffusive and the ballistic limit, the calculations are carried out at $T=1$~K. It is clear that the conductivity, the conductance, and the Seebeck coefficient show qualitatively similar behaviors in the diffusive and in the ballistic limit. We note that in both limits the Mott formula\cite{Cutler1969} gives a reasonable approximation of $S$ (not shown). The relatively larger conductivity in the electron regime in the diffusive limit is a result of the small effective mass associated with the CB, which gives a large mobility in the diffusive limit but has no effect on the conductance in the ballistic limit.

In general, an energy gap can be identified from the conductivity or the conductance by observing where these quantities drop to zero. The conductivities and conductances in Figs.~\ref{trans_GaSb}(a) and~\ref{trans_GaSb}(c) are finite for all energies, which confirms that these structures are metallic. For both of the gapped structures, cf. Figs.~\ref{GaSbInAs_Hybr}(b) and~\ref{GaSbInAs_Hybr}(d), we can similarly confirm the size of the energy gaps by examining the zero conductivities and conductances in Figs.~\ref{trans_GaSb}(b) and~\ref{trans_GaSb}(d).
An interesting feature of the Seebeck coefficient is that its sign reveals the transport characteristics of the system; a positive (negative) Seebeck coefficient indicates hole (electron) dominated carrier transport.

Based on the conductance trace in Fig.~\ref{trans_GaSb}(a), it would be difficult to tell whether we have an electron and/or a hole conductor, especially in the ballistic case. However, one sees that $S$ in Fig.~\ref{trans_GaSb}(e) has both positive and negative peaks for $E_F \lesssim 140$~meV, indicating that both electron- and holelike states contribute to the transport. The absence of hole-like transport, positive peaks in $S$, for $E_F \gtrsim 140$~meV corresponds to the top of the last VB-like band in Fig.~\ref{GaSbInAs_Hybr}(a). 

Figures~\ref{trans_GaSb}(b) and~\ref{trans_GaSb}(d) show the conductivities and conductances corresponding to the gapped nanowires in Figs.~\ref{GaSbInAs_Hybr}(b) and~\ref{GaSbInAs_Hybr}(d). Even though there is no qualitative difference between these conductivities, $S$ in Figs.~\ref{trans_GaSb}(f) and~\ref{trans_GaSb}(h) differ.
In general, $S$ for a gapped system always exhibits a large positive peak followed by a large negative peak, indicating hole- (electron)-dominated transport below (above) the band gap.
In Fig.~\ref{trans_GaSb}(f), we see that $S$ for the nanowire with a hybridization gap has an additional small negative peak just below the band gap (most clearly seen in the ballistic case). In contrast, in Fig.~\ref{trans_GaSb}(h), we see that $S$ for the nanowire with a confinement band gap lacks such a negative peak.
This negative peak in $S$, present only for nanowires with a hybridization gap, is even more profound for thinner core-shell nanowires (not shown).

Figure~\ref{trans_InAs} presents the calculated transport properties for a nanowire with an InAs core and a GaSb shell. For such wires, $\sigma$ and $G$ are less similar than for the inverted structure. From $S$ in Fig.~\ref{trans_InAs}(e), we can clearly see that that the VB-like bands end at $E_F \approx 65$~meV. For $E_F \gtrsim 65$~meV, we see that $S \approx 0$, corresponding to the lack of CB minima. In Fig.~\ref{trans_InAs}(f), it is possible to see where the CB-like bands end at just above $20$~meV, from studying the sign of $S$. This feature in $S$ cannot be seen in Figs.~\ref{trans_GaSb}(e)--(h) or Fig.~\ref{trans_InAs}(e) simply because the CBs end at a lower energy than what is visible in the plots. In Figs.~\ref{trans_InAs}(g) and~\ref{trans_InAs}(h), the appearance of $S$ is typical for a gapped system, with a large positive and a large negative peak around the band gap opening.

\section{Conclusions}
We have used $\mathbf{k\cdot p}$ theory to determine the band structures for $[111]$ GaSb/InAs and InAs/GaSb core-shell nanowires. The band structures show that by only varying the nanowire shell thickness, the system can be tuned between metallic and semiconducting, with a tunable band gap for the insulating system. For larger core radii, $R_C \gtrsim 20$~nm, a hybridization gap is present only in GaSb/InAs core-shell nanowires, and not in nanowires with an InAs core. We have used the energy dispersions as input to the Boltzmann equation and to ballistic transport equations to obtain the conductivity and the Seebeck coefficient.  The calculated transport properties are similar in both limits, especially for the case with nanowires with GaSb cores. Our results suggest that several band structure features can be seen from the Seebeck coefficient: the highest energy for where we have VB-like states, the lowest energy for where we have CB-like states, and the presence or absence of a hybridization gap.

\begin{acknowledgments}
The authors would like to thank Stephanie Reimann, Mats-Erik Pistol and Craig Pryor for fruitful discussions. 
This work was supported by the Swedish Research Council (VR), the Crafoord Foundation, the Ministry of Science and Technology of China (MOST) through the National Key Research and Development Program of China (Grant No. 2016YFA0300601), and the National Natural Science Foundation of China (Grants No. 91221202 and No. 91421303). Computational resources were provided by the Swedish National Infrastructure for Computing (SNIC) through Lunarc, the Center for Scientific and Technical Computing at Lund University.

\end{acknowledgments}

\appendix
\section{\label{app:kp}Parameters used in the $\mathbf{k\cdot p}$ calculations }
The parameters used in the eight-band Kane model are given by \cite{Foreman1997, Birner2011, Winkler2003,Willatzen2009}
\begin{align}
A &= \frac{\hbar^2}{2m_0}(1+2F) \\
P &= \sqrt{\frac{\hbar^2}{2m_0}E_P} \\
L &= -\frac{\hbar^2}{2m_0}(\gamma_1+4\gamma_2) + \frac{P^2}{E_g} \\
M &= -\frac{\hbar^2}{2m_0}(\gamma_1-2\gamma_2) \\
N &= -6\frac{\hbar^2}{2m_0}\gamma_3 + \frac{P^2}{E_g}
\end{align}
with the Kane energy $E_P$, the Luttinger parameters $\gamma_1$, $\gamma_2$, and $\gamma_3$, and the band gap $E_g$.  We follow the convention and set $F = 0$ to neglect the remote band contribution to the CB\cite{Bahder1990} in InAs. To eliminate possible spurious solutions, we follow the prescription in Ref.~\onlinecite{Foreman1997} by setting $A = 0$ and modifying $E_P$ for GaSb according to 
\begin{equation}
E_P = \frac{3m_0/m_c}{2/E_g + 1/(E_g + \Delta_{SO})}
\end{equation}
using the effective electron mass $m_c$. 
We use the symmetrization $N^+ = N^- = N/2$, since we have found that for the material system we are solving for, the difference between the symmetrization approach and the Burt-Foreman operator ordering results in a trivially small difference in the [001] direction. The magnitude of this difference is independent of crystal direction,\cite{Lassen2004} so  we can use the symmetrization scheme for a GaSb/InAs core-shell nanowire grown in the [111] direction (or any growth direction) without loss of precision. If this symmetrization scheme for $N$ is imposed, the rotation of the system from the [001] to the [111] crystallographic direction simplifies significantly.

The numerical values for the Luttinger parameters, band gaps, split-off band offsets $\Delta_{SO}$ and effective electron masses for GaSb and InAs are taken from Ref.~\onlinecite{Vurgaftman2001}, and can be found in Table~\ref{tab:params}. 
\begin{table}
\caption{\label{tab:params}Numerical values for the parameters used in the $\mathbf{k\cdot p}$ calculations}
\begin{ruledtabular}
\begin{tabular}{l|ll}
    & GaSb & InAs\\ \hline
 $\gamma_1$ & 13.4 & 20.0  \\
 $\gamma_2$ & 4.7 & 8.5  \\
 $\gamma_3$ & 6.0 & 9.2  \\
 $E_g$ & 0.812 eV& 0.417 eV \\
 $\Delta_{SO}$ & 0.76 eV & 0.39 eV\\
 $m_c$ & 0.039$m_0$ & 0.026$m_0$ \\
  $E_P$ & 24.82 eV & 21.5 eV \\

\end{tabular}
\end{ruledtabular}
\end{table}
The band offset between the VBs of the two different materials is given by $\Delta E = 0.56$~eV.\cite{Ekpunobi2005,Claessen1986,Sai-Halasz1978}
In the literature, a range of parameters extracted from experiments and first-principles calculations exist, and the parameters used in $\mathbf{k \cdot p}$ simulations are, many times, altered to avoid spurious solutions.

\section{\label{app:rot}Rotation of the $\mathbf{k\cdot p}$ Hamiltonian}
For a nanowire grown in a different crystal direction than $[001]$, we need to change the basis for the Hamiltonian and rotate our coordinate system. We follow a common approach for this procedure, described in detail by several authors.\cite{Willatzen2009,Lassen2006, NWLuo} The resulting Hamiltonian for the $[111]$ direction is given by
\begin{equation}
\widetilde{H} = Q^*W^* H_0' W^{\dagger}Q^{\dagger}+ H_{SO},
\end{equation}
where $H_0'$ is the $8\times 8$-Hamiltonian defined in Eq.~(\ref{HwoSO}), but with rotated coordinates ($k_i \rightarrow k_i'$). The matrix $W$ is a rotation matrix, while the matrix $Q$ accounts for an additional change to the new basis,
\begin{align}
\begin{split}
\left\lbrace \right.&\ket{S \uparrow}, \ket{S \downarrow}, \ket{HH \uparrow},  \ket{LH \uparrow},\\
&\left. \ket{LH \downarrow}, \ket{HH \downarrow}, \ket{SO \uparrow}, \ket{SO \downarrow} \right\rbrace.
\end{split}
\end{align}
The spin-orbit matrix $H_{SO}$ is invariant under rotation, so that it can be taken out of the calculations, together with the diagonal, $k$-independent part of $H$. The rotational matrix $W$ is given by
\begin{equation}
W = A \hat{U}
\end{equation}
with $A$ defined as
\begin{equation}
A = \begin{pmatrix}
\mathrm{e}^{-i \frac{\phi}{2}}\cos \frac{\theta}{2} \openone_{4\times 4} & \mathrm{e}^{i \frac{\phi}{2}}\sin \frac{\theta}{2} \openone_{4\times 4} \\
-\mathrm{e}^{-i \frac{\phi}{2}}\sin \frac{\theta}{2} \openone_{4\times 4} & \mathrm{e}^{i \frac{\phi}{2}}\cos \frac{\theta}{2} \openone_{4\times 4}
\end{pmatrix}
\end{equation}
and $\hat{U}$ as
\begin{equation}
\hat{U} =
\begin{pmatrix}
1 & 0 & 0 & 0  \\
0 & U & 0 & 0 \\
0 & 0 & 1 & 0 \\
0 & 0 & 0 & U
\end{pmatrix}
\end{equation}
with 
\begin{equation}
U = \begin{pmatrix}
\cos{\phi}\cos{\theta} & \sin{\phi}\cos{\theta} & -\sin{\theta} \\
-\sin{\phi}				& \cos{\phi}				& 0 \\
\cos{\phi}\sin{\theta} & \sin{\phi}\sin{\theta}  &\cos{\theta}
\end{pmatrix}.
\end{equation}
The angles $\phi$ and $\theta$ are defined as the azimuthal and the polar angles respectively, by which we rotate our coordinate system. 
The rotated Hamiltonian for the $[111]$ direction (with $N$ symmetrized and $B = 0$) is given by
\begin{equation}
\widetilde{H} = \left(\begin{array}{c|c|c}
H_{CC}	&	H_{CV8}	&	H_{CV7}	\\
\hline
\dagger	&	H_{V8V8}	&	H_{V8V7}	\\
\hline
\dagger		&	\dagger	&	H_{V7V7}
\end{array}\right).	\label{eq_Hrot}
\end{equation}
The submatrices of this rotated Hamiltonian are given by
\begin{equation}
H_{CC} = (E_c + k_xAk_x+k_yAk_y+k_zAk_z) \openone_{2\times 2},
\end{equation}
\begin{equation}
H_{CV8} = \begin{pmatrix}
i\frac{1}{\sqrt{2}}Pk_+	&	-i\sqrt{\frac{2}{3}}Pk_z		&	-i\frac{1}{\sqrt{6}}Pk_-		&	0 \\
0						& i\frac{1}{\sqrt{6}}Pk_+	&		-i\sqrt{\frac{2}{3}}Pk_z		&	i\frac{1}{\sqrt{2}}Pk_-
\end{pmatrix},
\end{equation}
\begin{equation}
H_{CV7} = \begin{pmatrix}
i\frac{1}{\sqrt{3}}Pk_z			&	-i\frac{1}{\sqrt{3}}Pk_-	 \\
i\frac{1}{\sqrt{3}}Pk_+			&  i\frac{1}{\sqrt{3}}Pk_z	
\end{pmatrix},
\end{equation}
\begin{widetext}
\begin{equation}
\begin{aligned}
&H_{V8V8} = \\
&\frac{1}{6}\begin{pmatrix}
N(\mathbf{k}^2 -3k_z^2)	 &	\frac{1}{\sqrt{3}}[\sqrt{2}(\Gamma_2-N)k_+^2-(2\Gamma_2+N)[k_-,k_z]_+] & \frac{1}{\sqrt{3}}[-(\Gamma_2 + 2N)k_-^2 + \sqrt{2}(\Gamma_2-N)[k_+,k_z]_+] &0\\
\dagger &				-N(\mathbf{k}^2 -3k_z^2)													& 0 &	-H_{V8V8}(1,3) \\
\dagger &					\dagger &	-N(\mathbf{k}^2 -3k_z^2)			& H_{V8V8}(1,2)	\\
 \dagger &					\dagger &	\dagger &	N(\mathbf{k}^2 -3k_z^2)
\end{pmatrix}\\
&+\frac{1}{6}(6E_v + 2\Gamma_1 \mathbf{k}^2)\openone_{4\times 4}
\end{aligned},
\end{equation}
\begin{equation}
H_{V8V7} = \begin{pmatrix}
\frac{1}{12\sqrt{3}}\left[-2(\Gamma_2-N)k_+^2 + \sqrt{2}(2\Gamma_2 + N)[k_-,k_z]_+\right] & \frac{1}{3\sqrt{6}}\left[-(\Gamma_2+2N)k_-^2 + \sqrt{2}(\Gamma_2 - N)[k_+,k_z]_+\right] \\
\frac{N}{3\sqrt{2}}(\mathbf{k}^2 -3k_z^2)	& \sqrt{3}H_{V8V7}(1,1) \\
\frac{1}{6}\left[(\Gamma_2 - N) k_-^2-\frac{1}{\sqrt{2}}(2\Gamma_2+N)[k_+,k_z]_+ \right]& H_{V8V7}(2,1) \\
\frac{1}{3\sqrt{6}}\left[(\Gamma_2 + 2N)k_+^2 - \sqrt{2}(\Gamma_2-N)[k_-,k_z]_+\right] & -\frac{1}{\sqrt{3}}H_{V8V7}(3,1) 
\end{pmatrix}
\end{equation}
\end{widetext}
and
\begin{equation}
H_{V7V7} = \frac{1}{3}(3E_v -3\Delta + \Gamma_1\mathbf{k}^2) \openone_{2\times 2}.
\end{equation}
Here, we use the substitution $k_{\pm}=k_x \pm i k_y$. Terms of the form $Dk_i^2$ ($Dk_ik_j$), where $D$ is any parameter, are to be interpreted as $k_iDk_i$ ($k_iDk_j$) and $\mathbf{k}^2 = k_x^2 + k_y^2 + k_z^2$. The anti-commutation notation $[k_i,k_j]_+ = k_ik_j + k_jk_i$ is employed, and the substitutions
\begin{align}
&\Gamma_1 = L+2M \\
&\Gamma_2 = L-M
\end{align}
are used throughout.

\bibliography{library_modified}

\end{document}